\documentclass[traditabstract]{aa}

\usepackage{graphicx}
\usepackage{txfonts}
\usepackage{multirow}
\usepackage{longtable}
\usepackage{color}

\begin{document}

\newcommand{\kms}{km\,s$^{-1}$}
\newcommand{\um}{\mu}
\title{Post-outburst spectra of a stellar-merger remnant of V1309\,Scorpii:\\ 
       from a twin of V838\,Monocerotis to a clone of V4332\,Sagittarii}

\author{T. Kami\'nski\inst{\ref{inst1},\ref{inst2}}, 
        E. Mason\inst{\ref{inst3}}, 
        R. Tylenda\inst{\ref{inst4}}
   \and M. R. Schmidt\inst{\ref{inst4}}
        }

  \institute{\centering ESO, Alonso de C\'ordova 3107, Vitacura, Casilla 19001, Santiago, Chile, \email{tkaminsk@eso.org}\label{inst1}
       \and Max-Planck-Institut f\"ur Radioastronomie, Auf dem H\"ugel 69, 53121 Bonn, Germany,\label{inst2} 
       \and  INAF Osservatorio Astronomico di Trieste, via G. B. Tiepolo 11, 34134, Trieste, Italy\label{inst3}
       \and  Nicolaus Copernicus Astronomical Center, Polish Academy of Sciences, Rabia\'nska 8, 87-100 Toru\'n\label{inst4}
       }
            
  \date{Received; accepted}

\abstract{We present optical and infrared spectroscopy of V1309\,Sco, an object that erupted in 2008 in a stellar-merger event. During the outburst, V1309\,Sco displayed characteristics typical of red transients, a class of objects similar to V838\,Mon. Our observations were obtained in 2009 and 2012, i.e. months and years after the eruption of V1309\,Sco, and illustrate severe changes in the remnant, mainly in its circumstellar surroundings. In addition to atomic gas observed in earlier epochs, we identified molecular bands of TiO, VO, H$_2$O, ScO, AlO, and CrO. The infrared bands of CrO we analyse are the first astronomical identification of the features. Over the whole period covered by our data, the remnant was associated with a cool ($\lesssim$1000\,K) outflow with a terminal velocity of about 200\,\kms. Signatures of warmer atomic gas, likely to be still dissipating the energy of the 2008 outburst, dramatically decreased their brightness between 2009 and 2012. In addition,  the source of optical continuum disappeared sometime before 2012, likely owing to the formation of new dust. The final stage of V1309\,Sco's evolution captured by our spectra is an object remarkably similar to an older red transient, V4332\,Sgr. In addition to providing a detailed view on the settling of the eruptive object, the observations presented here reinforce the conclusion that all the Galactic red transients are a manifestation of the same phenomenon, i.e. a stellar merger. The late spectra of V1309\,Sco also suggest peculiarities in the chemical composition of the remnant, which still need to be explored.}
 

\authorrunning{Kami\'{n}ski et al.}
\titlerunning{Stellar merger remnant V1309\,Sco}
\maketitle

\section{Introduction}
V1309\,Sco was discovered in outburst in 2008 by K. Nishiyama and F. Kabashima in Japan and independently by Guoyou Sun and Xing Gao in China \citep{nakano}; it was initially classified as a nova (Nova Scorpii 2008). Detailed spectroscopic observations of the object during the outburst and in its early decline showed, however, that it did not follow the  behaviour of classical novae \citep{mason}. Low outflow velocities of a few hundred \kms, a presence of low-excitation gas, and -- most importantly -- evolution towards cool phase characterized by an M-type continuum, pointed towards classification of V1309\,Sco as a `red nova' or a red transient (the group is also known as intermediate-luminosity optical transients or mergebursts). So far, the best studied object in this group is V838\,Mon, which erupted in 2002, and therefore the red transients are also known as V838-Mon-type objects. This newly recognized class of variables has been considered to erupt as a result of a stellar collision and consequent coalescence \citep{ST03,ST06,TS06,ST07}. The case of V1309\,Sco provided the best and direct observational evidence that red transients indeed erupt as a result of a stellar merger. \citet{tylenda} presented photometry of the object going back nearly seven years  before its outburst and showed that it was an eclipsing binary system with a period of about 1.4\,days. This period was shortening at an exponential rate before the 2008 eruption showing directly, for the first time, a spiralling-in process that eventually led to a merger. Given that V1309\,Sco showed observational characteristics typical of red transients and the strong evidence for the merger in 2008, the object is considered  the `Rosetta stone' linking the class of eruptive variables with mergers, solely on  strong observational grounds. Five galactic objects belong to this group, including OGLE-2002-BLG-360 \citep{blg}; the oldest of them are V4332\,Sgr, which erupted in 1994 \citep{martini,tylcrause}, and possibly CK\,Vul with a very ancient eruption in 1670 \citep{blg,CK}. Many of the recognized red transients have been observed in nearby galaxies, e.g. M31\,RV \citep{rich,mould}, M31N 2015-01a \citep{tomov,williams}, and M85\,OT2006-1 \citep{kulkarni}. Red transients (hence mergers) are now considered to be a more common phenomenon than thought a few years ago, with an average Galactic rate of 1 event per 2\,yr \citep{kochanek}. Many will be observed with the current and future sky surveys. 

Observing red novae is the best way to study the merger process, but  theoretical effort has also been undertaken \citep{nandez}.  After their outbursts all red transients develop complex circumstellar environments abundant in molecules and dust. Their characteristics are linked to the crucial information on the physics of a merger, for instance the mass and angular momentum lost from the system during the violent event. Here, we present spectra obtained with Xshooter on the Very Large Telescope (VLT) demonstrating spectral evolution of V1309\,Sco in 2009 and 2012, i.e. months to years after its spectacular outburst. The spectra cover the range between about 300\,nm and 2.48\,$\um$m providing a very comprehensive view on the post-merger remnant of V1309\,Sco.  This study is a continuation of spectroscopic monitoring of this object by \citet{mason}, whose last published observations were obtained on 21 April 2009, i.e. 227 days after the maximum light.  

\begin{table*}
 \centering
 \caption{Technical details of the observations.}
 \label{tab-log}
\begin{tabular}{lcc|ccc|ccc|ccc}
\hline
       & Days &Slit & \multicolumn{3}{c|}{UVB arm} & \multicolumn{3}{c|}{VIS arm} & \multicolumn{3}{c}{NIR arm} \\ 
\multicolumn{1}{c|}{UT date}& since&PA\tablefootmark{a}& exposure & slit & binning &exposure & slit & binning &exposure & slit & binning \\
       & max. &(deg)& time (s) & width (")  & (pix)   &time (s) & (")  & (pix)   &time (s) & (")  & (pix) \\
\hline
2009-May-04  & 240  &1.7& 600$\times$2 & 1.0 & 1$\times$1 & 600$\times$2 & 0.9 & 1$\times$1 & 200$\times$6 & 0.9 & 1$\times$1 \\
2009-May-04  & 240  &1.7& 900$\times$2 & 1.0 & 1$\times$1 & 400$\times$4 & 0.9 & 1$\times$1 & 450$\times$4 & 0.9 & 1$\times$1 \\[5pt]
2009-Aug-13 & 341&51& 1000$\times$2 & 1.0 & 1$\times$2 & 1000$\times$2 & 0.9 & 1$\times$2 & (300$\times$3\tablefootmark{b})$\times$2 & 0.9 & 1$\times$1\\
2009-Sep-29 &388 &51 & 1000$\times$2 & 1.0 & 1$\times$2 & 1000$\times$2 & 0.9 & 1$\times$2 & (300$\times$3\tablefootmark{b})$\times$2 & 0.9 & 1$\times$1\\[5pt]
2012-Apr-22 & 1324 &--72& 200 & 5.0 & 1$\times$1 & 200 & 5.0 & 1$\times$1 & 268 & 5.0 &  1$\times$1\\
2012-Apr-22 & 1324 &--72& 394$\times$2 & 1.0 & 1$\times$1 & 394$\times$2 & 0.7 & 1$\times$1 & 461$\times$2 & 0.4 & 1$\times$1 \\
\hline
\end{tabular}
\tablefoot{
\tablefoottext{a}{Position angle measured from North through East.}
\tablefoottext{b}{We did not obtain six separate exposures but rather set NDIT parameter \citep{XSHmanual} to 3, so that single frame was stored for each nodding position and corresponds to a pre-processor average of three integrations of 300\,s each. }
}
\end{table*}

\section{Observations}
All observations of V1309\,Sco were conducted with Xshooter \citep{xsh}, which is a spectrograph mounted on Unit Telescope 2 of VLT. Xshooter registers spectra in three fragments simultaneously: ultraviolet within 300--560\,nm, visual within 550--1020\,nm, and near-infrared between 1020 and 2480\,nm. We refer to the three spectral ranges as UVB, VIS, and NIR, respectively. 

All the spectra were acquired in the nodding mode to better correct the NIR data for the sky background. The length of the nodding throw was 5\arcsec. The position angle of the nodding direction was different for different dates. The slit length in all the observations was 11\arcsec. The observations were conducted with the atmospheric dispersion corrector inserted in the optical path. The detector was always used in the high-gain mode. A technical summary of the observations is provided in Table\,\ref{tab-log}. 

\paragraph{Commissioning data}
The first observations were obtained during one of commissioning runs of Xshooter\footnote{ESO programme 60.A-9024(A), see \url{https://www.eso.org/sci/activities/vltcomm/xshooter.html}}. The observations were obtained on 4 May 2009 and were accompanied by seeing of 0\farcs7. In addition to V1309\,Sco, the slit covered a nearby star $\sim$2\arcsec\ north from the eruptive object. The number of exposures varied for each arm (see Table\,\ref{tab-log}) and the exposure times between 200 and 900\,s resulted in the total time on source of about 50\,min. Even the shortest exposures in the VIS arm of 400\,s resulted in the saturation of the H$\alpha$ line. We used the slit widths of 1\arcsec\ for the UVB arm and of 0\farcs9 for the two other arms. This provided us with spectra at the nominal resolving power of $R\!=\!\lambda/\Delta\lambda$=5100, 8800, and 5300, respectively. 

The telluric standard star HIP89486 (spectral type B9\,V) was observed immediately after the science observations, while the spectrophotometric standard star EG274 was observed a few hours later at the end of the commissioning run. 

\paragraph{Science-verification data}
Later in 2009, additional observations were  obtained within project 60.A-9445(A) (PI: E. Mason), which was part of the science verification of Xshooter. V1309\,Sco was observed twice on 12 August and 29 September. Seeing during the observations was of 1\farcs0. The slit widths and resolution reached were the same as in the earlier commissioning data, but the detectors of the UVB and VIS arms were binned in the spatial direction (1$\times$2 pix). The slit was aligned with the parallactic angle at the beginning of the exposure. A series of exposures was taken (Table\,\ref{tab-log}) adding to the total integration time of about 33\,min both in August and September.

The telluric standard HIP088857 (spectral type B3\,V) was observed at the two science-verification runs. The spectrophotometric standard star BD+17\,4708 (spectral type sdF8), was observed only during the August run, but we used it to calibrate data from both runs. 

\paragraph{The 2012 data}
Another set of observations was obtained on 22 April 2012 within a service-mode program 089.D-0041(A) (PI: T. Kami\'nski). Two spectral setups were used. The first  employed the widest available slits of 5\arcsec\ to minimize slit losses and to provide a spectrum suitable for accurate flux calibration. The spectral resolution was between 1900 and 3200 in the three spectral ranges. 
The second setup was aimed to provide spectra at a higher resolution and a set of narrower slits was used: 1\farcs0, 0\farcs7, 0\farcs4 for UVB, VIS, and NIR, respectively. This provided us with spectra at nominal resolutions of $R$=4\,300 for UVB, and of 11\,000 for VIS and NIR. 
The average seeing during the 2012 observations was of 0\farcs5. The 2012 observations were considerably shorter than those in 2009 (cf.\,Table\,\ref{tab-log}).


During the 2012 observing run, the telluric standard HIP067796 (spectral type B2\,II) and the spectrophotometric standard EG274 (spectral type DA) were observed on the same night as V1309\,Sco. A telluric standard was observed with a slit width of 0\farcs9 in VIS and NIR arms, and the flux standard was observed with the wide 5\arcsec slits. 


\section{Data reduction}
The commissioning and science-verification data from 2009, as test observations, required  special care when basic image processing was applied. The primary data reduction was obtained using the {\tt esorex} ESO pipeline (version 1.2.2) and included flat-field correction, echelle-order rectification, and wavelength calibrations. For the commissioning data, all the necessary calibration frames were taken the day following the science observations. However, spectra of calibration lamps were underexposed and could not be used alone. We therefore used corresponding frames taken a few months later. We found the same strong emission lines in the underexposed and well-exposed spectra of calibration lamps; based on their comparison we can exclude systematic errors in the wavelength calibration due to possible (but unlikely) changes in the spectral format \citep{XSHmanual}. Further processing of all the 2009 data --- including the order extraction, order merging, telluric correction, and flux calibration --- was performed using standard procedures in IRAF. We applied the telluric correction in IRAF only to the NIR-arm spectra. 
The telluric correction in the VIS region was applied to the flux-calibrated spectra using {\tt molecfit} \citep{molecfit}. This tool corrects for the telluric absorption by performing a full radiative-transfer simulation of the atmospheric absorption bands at weather conditions specific for the science observations. The telluric correction was calibrated on the visual O$_2$ bands, which do not have any circumstellar components. The contribution of telluric water vapour was then calculated using a standard H$_2$O-to-O$_2$ abundance ratio. The applied procedure left residua below $\sim$10\% of the original telluric absorption. 


The basic data reduction of the 2012 data (including flat-field corrections, wavelength, and flux calibrations) were obtained using the standard Xshooter pipeline under the Reflex environment\footnote{\url{http://www.eso.org/sci/software/pipelines/}}. However, owing to a severe contamination by field stars, the spectral extraction had to be performed manually and separately for each of the nodding phases. This was done in IRAF. For reference, the 2012 data were fully reduced using the pipeline in different versions, and we also made use of the Xshooter Data Release 3\footnote{\url{http://www.eso.org/observing/dfo/quality/PHOENIX/XSHOOTER/processing.html}}, especially for the presentation of the NIR-arm data. The cumbersome spectral extraction owing to the field crowding and an unfortunate  position angle of the nodding direction practically exclude any reliable flux calibration of the 2012 data. Therefore, the data obtained with the wide slit are not presented here. We corrected the 2012 spectra for telluric absorption using {\tt molecfit}. Because the NIR spectrum has no significant continuum, the {\tt molecfit} correction procedure could not be applied for it.
 
The wavelength calibration of the science-verification and the 2012 data is accurate within an rms of 0.002\,nm in the UVB and VIS range, and 0.01--0.02\,nm for the NIR arm.  All spectra and radial velocities are presented here in the heliocentric rest frame, $V_h$. The correction from $V_h$ to the local standard of rest (LSR) velocity at the position of V1309\,Sco is $V_{\rm LSR}$=$V_h$+9.775\,\kms. 

\begin{figure}\centering
\includegraphics[angle=0,width=\columnwidth]{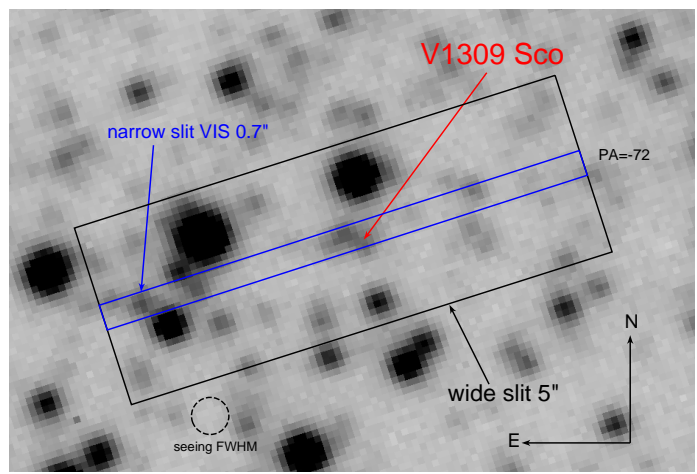}
\caption{ Location of the VIS slits in the 2012 observations.}\label{fig-slit}
\end{figure}

In Fig. \ref{fig-slit} we show the Xshooter acquisition image illustrating the location of the slits in the 2012 observing run. The image was obtained in the $V$ filter in 2012. The wide slit encompassed several bright stars in the field, in particular the star 2\arcsec\ north from the position of V1309\,Sco. This source dominates or is comparable to continuum brightness of V1309\,Sco from optical to NIR wavelengths in the time period covered by our observations (Kami\'nski et al. in prep.). Images obtained with the Hubble Space Telescope \citep{mccollum} also resolve  the source seen in the Xshooter image at the position of V1309\,Sco into at least three separate sources, so also the observations with the narrow slits gave spectra considerably affected by the field stars. An inspection of our earliest-epoch data, i.e. when V1309\,Sco was brightest, show that even then the spectra were contaminated by the field stars. Because of this blending effect and the rough absolute-flux calibration, we focus here on an analysis of spectral features in the spectrum of V1309\,Sco, not the continuum. In particular, we \emph{\emph{subtracted}} the false continuum of the 2012 spectrum as a low-order polynomial. We \emph{\emph{did not}} correct the spectra for interstellar extinction.

\section{Identification and analysis of spectral features}
The VIS and NIR spectra with the proposed identification are shown in Figs.\,\ref{fig-vis-id}--\ref{fig-nir-B}. (The UVB spectrum will be presented elsewhere.) To identify atomic emission lines, we used  the database of NIST \citep[National Institute of Standards and Technology;][]{nist}. The location and structure of molecular bands was simulated mainly in {\tt pgopher} \citep{pgopher} using spectroscopic constants and literature data described in detail in Appendix\,\ref{appen}. 

As shown in Fig.\,\ref{fig-dates}, the spectra were obtained during the fading part of  V1309\,Sco's light curve  and the continuum flux changed by a factor of about 24 from the earliest to the latest spectrum. We analyse the spectra emphasizing the time variability of spectral features in that period.

\begin{figure}\centering
\includegraphics[angle=270,width=\columnwidth]{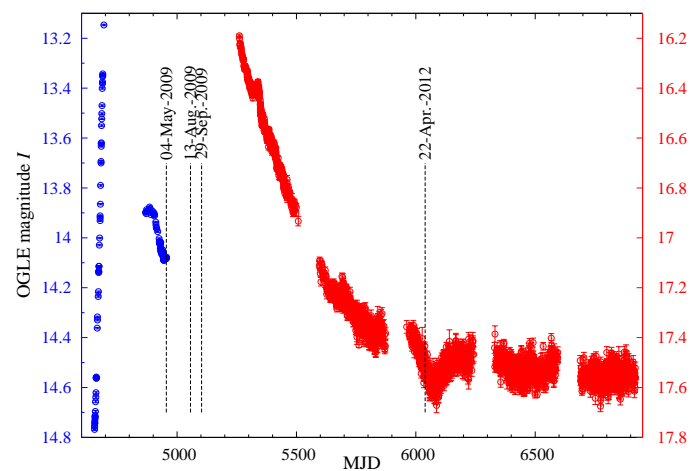}
\caption{Dates of Xshooter observations marked on the light curve of V1309\,Sco as observed by OGLE in the $I$ filter \citep{tylenda}. The light curve before and after MJD=5000 is shown in different scales as indicated on the left (blue) and right (red) axes.}\label{fig-dates}
\end{figure}

\subsection{UVB range}

Only very weak continuum rising towards longer wavelengths was detected in the UVB range. Furthermore, weak absorption features are present longward of about 490\,nm, with the \ion{Mg}{I} triplet (Mg\,b, 516.7--518.4\,nm) and H$\beta$ being the most prominent examples (the former seen clearly only in the 2012 spectrum). The continuum and absorption spectrum most likely belong to the field star(s) of spectral types close to G--K. All the UVB spectra are very rich in emission lines, which undoubtedly do belong to V1309\,Sco. 

Each spectrum acquired in 2009 has over 60 spectral features. Although obtained weeks and months apart during the very steep dimming part of  V1309\,Sco's light curve (Fig.\,\ref{fig-dates}), the 2009 spectra appear very similar to each other when scaled to the same intensity. No spectral features have been found above the noise levels shortward from about 380\,nm, except for the earliest Xshooter spectrum where emission lines are seen down to at least 320\,nm. 

In the 2012 spectrum, which has the lowest signal-to-noise ratio, only the strongest of the lines seen in 2009 are clearly visible and many of them are narrower in 2012 (all UVB spectra have a very similar resolution). A few new features are present in the latest spectrum, e.g. near 480.5 and 490.6\,nm. 

In all the spectra,  the most prominent emission lines are the Balmer features (except for the  2012 spectrum where they are absent), individual lines of [\ion{O}{I}] and \ion{Mg}{I}], and two lines of \ion{S}{II}. In addition, several multiplets of \ion{Fe}{I} and \ion{Fe}{II} (allowed and forbidden) and the H\&K lines of \ion{Ca}{II} were securely identified in the spectra. However, more than half of the observed spectral features could not be unambiguously identified. Many of those unidentified features are at wavelengths of known lines of Fe, Ti, Cr, V, Ni, and their ions, but not all lines in the given multiplet could be traced in the spectrum, hampering a definite  assignment. Judging from the relative intensities of the identified lines of \ion{Fe}{I} and \ion{Fe}{II}, the emission spectrum does not arise from  istothermal gas. Many lines may be produced by scattering which results in line ratios strongly deviating from those expected from emission of gas in thermal equilibrium. 

When  the emission spectra from the different epochs are compared in absolute scale, a clear decrease in line fluxes with time is apparent. The continuum-free integrated flux in the 385--559\,nm range dropped from 13.9 (May 2009) to 2.7 (August 2009) and then to 1.9 (September 2009) in units of 10$^{-15}$\,erg\,cm$^{-2}$\,s$^{-1}$. The same measurement for the 2012 spectrum, although with very uncertain flux calibration, gives 6.3$\times$10$^{-17}$\,erg\,cm$^{-2}$\,s$^{-1}$. The decrease in the average line fluxes is therefore steeper than the corresponding drop in $I$-band continuum documented by the OGLE light curve (Fig.\,\ref{fig-dates}). The different lines fade in a slightly different rate. In particular, the \ion{H}{I} lines fade much faster than those of all the observed metals; for H$\beta$, the measured flux in units of 10$^{-15}$\,erg\,cm$^{-2}$\,s$^{-1}$ goes from 39.3 (May) to 2.7 (August) and to 2.2 (September) in 2009 spectra, and the line is absent in the 2012 spectrum. The \ion{Ca}{II} H\&K lines faded very quickly as well.

Although spectral features are present close to the predicted positions of heads of the $B-X$ band of AlO (e.g. near 485\,nm and 507.5\,nm) and the $\alpha$ (0,0) band of TiO (at 516.5\,nm), there is no convincing evidence for a presence of molecular emission bands in the UVB range. 


\subsection{VIS range}\label{sect-vis}
\subsubsection{Atomic lines}
In the VIS range of the four Xshooter spectra, we found only a few atomic lines, among which the strongest are the H$\alpha$ line; lines of [\ion{O}{I}] (630.0 \& 636.4\,nm) and [\ion{Fe}{II}] (e.g. 544.0, 564.0, 569.6\,nm), the resonance doublets of \ion{K}{I} and \ion{Rb}{I}; the semi-forbidden line of \ion{Ca}{I} (seen outside the H$\alpha$ line only in 2012); and the resonance pair of \ion{Cs}{I} lines (684.9 \& 689.5\,nm), absent in 2012. 

\begin{figure*}\centering
\includegraphics[angle=270,width=\textwidth]{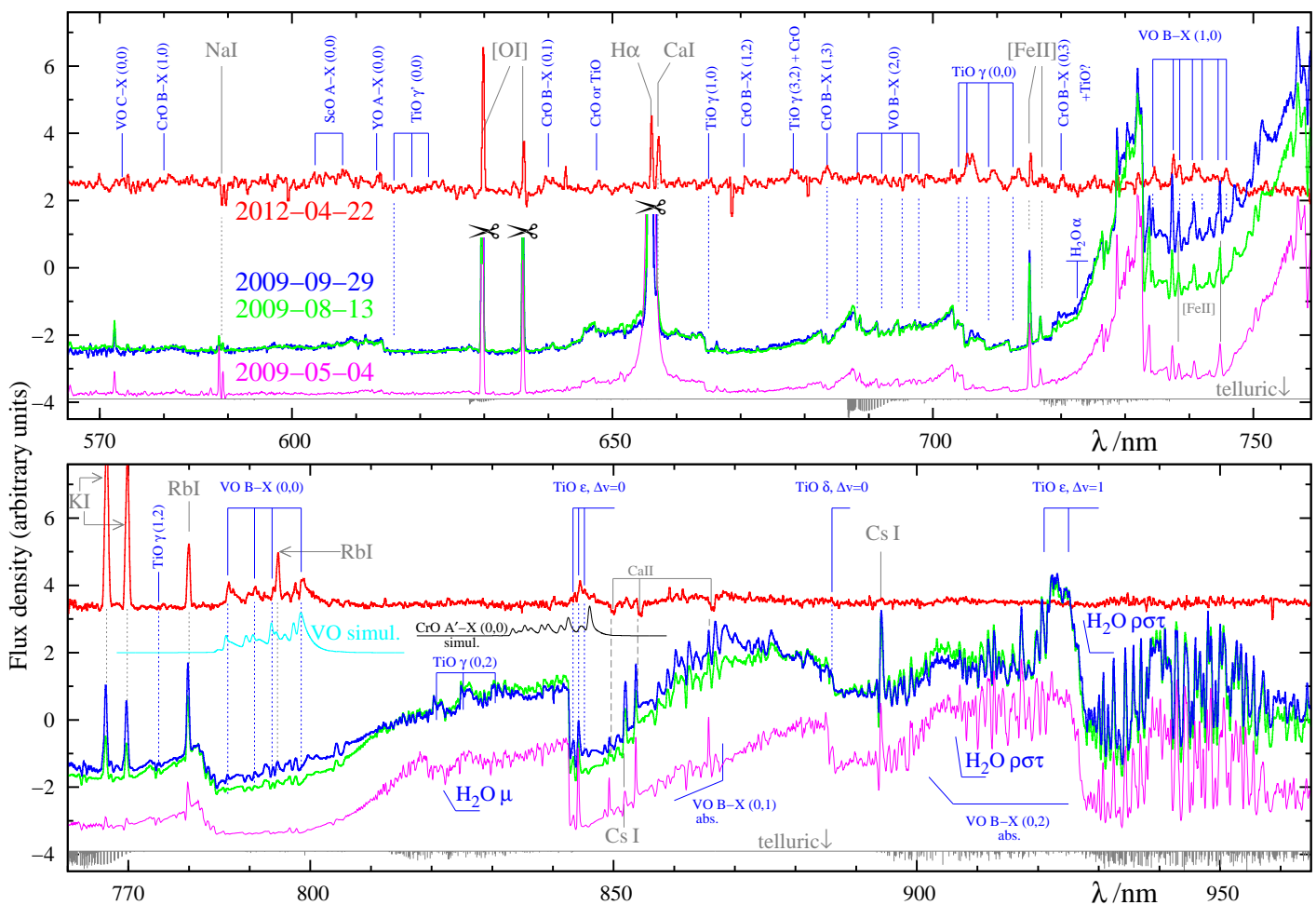}
\caption{Identification of spectral features in the VIS range. The spectra from May 2009, August 2009, September 2009, and April 2012 are shown with magenta, green, blue, and red lines, respectively. 
The spectra were corrected for telluric absorption, but location and structure of telluric bands is shown at the bottom part of each panel (grey line). A continuum was subtracted from the 2012 spectrum. All spectra were smoothed and are not in scale with respect to each other. The top parts of the profiles of the H$\alpha$ and the [\ion{O}{I}] lines in the 2009 spectra were clipped for clarity and are marked with the scissors (some profiles are compared separately in Fig.\,\ref{fig-profiles}). Simulations of the (0,0) band of the VO $B-X$ and CrO $A'-X$ systems are shown in the lower panel below the 2012 spectrum (in cyan and black, respectively). They were generated with {\tt pgopher} at an excitation temperature of 300\,K (see Appendix\,\ref{appen}).}\label{fig-vis-id}
\end{figure*}

While the 2009 spectra are undoubtedly dominated by the emission and absorption features of V1309\,Sco itself, the spectrum obtained in 2012 -- when the object was substantially weaker in continuum ($\sim$20 times in $I$) -- is contaminated by spectra of the blending objects that have a comparable brightness to V1309\,Sco (Fig.\,\ref{fig-slit}). The 2012 spectrum contains narrow absorption features of which the \ion{Na}{I} doublet at 589\,nm and the \ion{Ca}{II} triplet near 860\,nm are most clearly recognizable. These absorption features most likely come from the blending stars whose spectral type must be close to G--K, considering the strengths of the mentioned lines. At a closer inspection, the \ion{Ca}{II} triplet also appears to have  weak and blue-shifted (with respect to the absorption line centre) emission component, which can be ascribed to V1309\,Sco. This is supported by the presence of pure emission lines of \ion{Ca}{II} in the earlier epochs. 

The time evolution of most emission lines is linked to what was described above for lines in the UVB range -- most of them quickly fade with time. These include H$\alpha$ and features of \ion{Fe}{II}, \ion{Ca}{II}, and [\ion{O}{I}]. This group of lines gets fainter with time in terms of absolute flux units and also with respect to the continuum. 

An interesting behaviour is observed in resonance lines of alkali metals. The line of \ion{Cs}{I} near 894.3\,nm gets weaker with time over 2009 to completely disappear in 2012. 
The $\lambda$7800 line of \ion{Rb}{I} in 2009 was getting fainter with time in absolute flux units, but was gradually gaining its strength with respect to the local continuum (Fig.\,\ref{fig-vis-id}). In the 2012 spectrum, which however has very poor flux calibration, the $\lambda$7800 line is comparable or even higher than in the last spectrum from 2009. The other line of the \ion{Rb}{I} doublet, $\lambda$7944, although clearly present in 2012, is not seen at all in the 2009 spectra, possibly because it is absorbed by the saturated absorption band of VO. This implies, that the source of atomic emission must be internal to (behind) the cloud of absorbing molecular gas. The optical doublet of \ion{K}{I} shows very similar behaviour to the $\lambda$7800 line of \ion{Rb}{I} -- over the 2009 observations, it fades on a pace lower than the decay of the underlying continuum but in 2012 the doublet is clearly stronger than in any of the 2009 spectra. 

The optical doublet of \ion{Na}{I} shows a different behaviour. In 2009 spectra, the lines show weak net emission despite being absorbed by interstellar absorption components \citep[cf.][]{mason} and photospheric lines of the blending field stars. The emission gradually fades in time during 2009 observations. Only very weak emission is seen in the 2012 spectrum, comparable to that observed in the first 2009 spectrum. Because the \ion{Na}{I} usually produces stronger lines than \ion{K}{I} in a wide range of astrophysical environments and physical conditions, the observed characteristics of those two doublets in V1309\,Sco is rather peculiar. A similar peculiarity has been noticed in V4332\,Sgr and is explained as the result of the scattering of light of a source that is   hidden by dust  \citep{kami4332,tylUVES}. This feature bears crucial information on the emission mechanism and source geometry as discussed in more detail in Sect.\,\ref{sect-discuss}.

\subsubsection{Molecular bands and pseudo-continuum}
\begin{figure}\centering
\includegraphics[angle=270,width=\columnwidth]{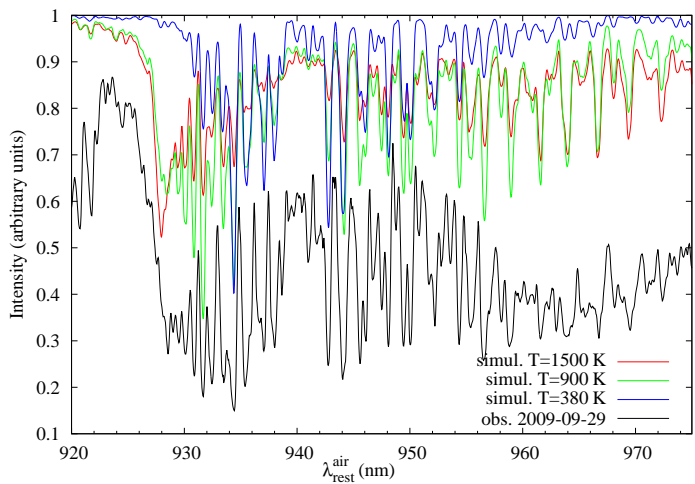}
\caption{Water absorption band in V1309\,Sco (black) is compared to simulations of the $\rho\sigma\tau$ band at different excitation temperatures (see Appendix\,\ref{appen}). The observed band cannot be formed as a telluric feature because the structure of the band at air temperatures does not form a clear band-head near 926\,nm. All simulations were smoothed to fit the width of the observed features. Although water dominates, there are additional opacity sources in this wavelenght range, e.g. the TiO $\epsilon$ system.}\label{fig-water}
\end{figure}

The VIS range is abundant in molecular bands.
The 2009 spectra are very similar to each other and show a cool continuum absorbed in bands of simple oxides which are typical for late M-type stellar photospheres. These bands belong mainly to the electronic systems of TiO ($\gamma$, $\gamma^{\prime}$, $\epsilon$, $\delta$) and VO ($B-X$) (Fig.\,\ref{fig-vis-id}). Additionally, above 890\,nm, we observe strong absorption of water, most clearly in the $\rho\sigma\tau$ ro-vibrational band \citep{waterVIS} whose upper level is 7650\,K above the ground. The weaker visual bands of water, $\alpha$ ($\sim$723\,nm) and $\mu$ ($\sim$822\,nm), are also present but less conspicuous. We are confident the absorption in the $\rho\sigma\tau$ bands is intrinsic to V1309\,Sco and not to the Earth's atmosphere because the lines are significantly broader than the instrumental profile and are shifted to $V_{\rm h}$ of $\sim$--160$\pm$20\,\kms. Additionally, we compared the observed spectrum (corrected for the telluric features) to a grid of simulations of the $\rho\sigma\tau$ band at different excitation temperatures. The observed shape of the head near 927\,nm can  only be formed in excitation temperatures significantly higher than those in the Earth's atmosphere, as illustrated in Fig.\,\ref{fig-water}. At temperatures $\lesssim$380\,K strong abortion is seen only above 932\,nm, i.e. 6\,nm longward from the observed head. The location of the band-head can be very well reproduced for 850\,K$>T_{\rm ex}\!>$1000\,K; this can be treated as the first rough constraint on the temperature of the absorbing circumstellar vapour. There are other opacity sources overlapping with the water band, but -- owing to the dominant contribution of water -- they could not be conclusively identified. In particular, emission-like features at 966.9, 976.4, 989.9, and 1004.6\,nm may be related to the vibrational band of OH (Fig.\,\ref{fig-nir}). 

With respect to the continuum, the absorption bands were deepest in the earliest observation of May 2009 and were considerably weakened in the August/September 2009 observations. In most of the bands, we observe consequent weakening of the bands depth with time. Despite the imperfect flux calibration, we are confident that this effect is real and not related to an imperfect correction for the instrument response function. For instance, the shape of the absorption through of the VO $B-X$ (0,0)\footnote{To assign the rovibrational bands, we use the convention ($\varv_{\rm up},\varv_{\rm low}$) where the numbers are the upper and lower vibrational states.} band (in 760--810\,nm) changes with time as expected for consequent lowering of the column density of absorbing gas --- there are almost no changes near the band-head while  significant changes are seen in the tail of the  band  where opacity is lowest.   

We made an attempt to derive the spectral type for the 2009 observations assuming the pseudo-continuum is produced by a stationary stellar-like photosphere. By comparing the observations to simulated spectra of the MARCS stellar-atmosphere models \citep{marcs}, however, we found that the shape of the pseudo-continuum strongly deviates from any M-type photosphere with a standard chemical composition (i.e. close to solar). Most apparent in V1309\,Sco is the greatly enhanced absorption within the $B-X$ system of VO and the $\rho\sigma\tau$ band of H$_2$O. A similar effect in VO bands has been reported for the spectrum of V4332\,Sgr \citep{kami4332}, while strong optical absorption of water was seen in another red transient V838\,Mon \citep{lynch1,lynch2}. Such strong absorption indicates the presence of cool circumstellar gas surrounding the stellar remnant. Although we compared the spectrum to a broad range of photospheric models applying different values of reddening, we were not able to assign a spectral type in the classical sense. Even if the spectrum arises in a stellar-like object, its composition may be very different from a normal star and spectral typing would then inevitably fail \citep[cf.][]{kami4332}. Nevertheless, the overall slope of the observed spectrum is close to standard spectra generated for effective temperatures $\lesssim$3300\,K or spectral types later than M5. Additionally, it is clear that the spectrum is formed in an extended atmosphere, i.e. of a giant or a supergiant, because absorption features of simple hydrates (e.g. FeH, MgH, CaH, CaOH, CrH) -- usually present in spectra of dwarfs -- are absent. 

Some molecular bands also showed  weak \emph{\emph{emission}} in the 2009 spectra. The $B-X$ system of VO is dominated by absorption, but  within the absorption dips  weak emission features of VO are also seen, most clearly in the (1,0) band near 740\,nm (Fig.\,\ref{fig-vis-id}). 

The molecular absorption bands disappeared almost completely in 2012, which is directly related to the dramatic (apparent)  fading of the continuum source (Fig.\,\ref{fig-dates}). Most of the molecular bands seen in absorption in 2009 turned into a pure emission spectrum in 2012 (excluding bands of water). In addition to the well-known electronic systems of TiO and VO, the 2012 emission spectrum contains weak but clear bands of CrO belonging to the $B-X$ system, most clearly (and unambiguously) in the (0,1) band near 640\,nm. This band has been observed in emission only in one other astrophysical source, V4332\,Sgr \citep{kami4332}. We also found evidence for weak emission of CrO in another electronic system, $A^{\prime}-X$, which overlaps with emission of the TiO $\epsilon$ (0,0) band near 843\,nm (Fig.\,\ref{fig-vis-id}). Because the bands overlap and have complex structure, this identification is only tentative at the moment; it is supported, however, by the presence of CrO emission in other electronic systems within the VIS and NIR spectra (see below). 

Other less common species seen in emission in 2012 are the $A\,^2\Pi_{3/2}-X\,^2\Sigma^+$ bands of ScO and YO (within 590--615\,nm, see Fig.\,\ref{fig-vis-id}), which are however weak. No features of circumstellar water vapour are seen in the 2012 spectrum above the residuals of the telluric correction. 

\subsection{NIR range}
Atomic features are relatively rare in the NIR range, with only recombination lines of hydrogen (e.g. Paschen H$\beta-\delta$, possibly Brackett H$\gamma$) having a reliable identification. A narrow emission feature at 1065.6\,nm seen in the two spectra from 2009 may be a transition of \ion{Ti}{I} or [\ion{Sc}{II}]. The NIR range is dominated by molecular features that are described in detailed below, separately for each of the identified species.

\subsubsection{AlO}
\begin{figure*}\centering
\includegraphics[width=\textwidth]{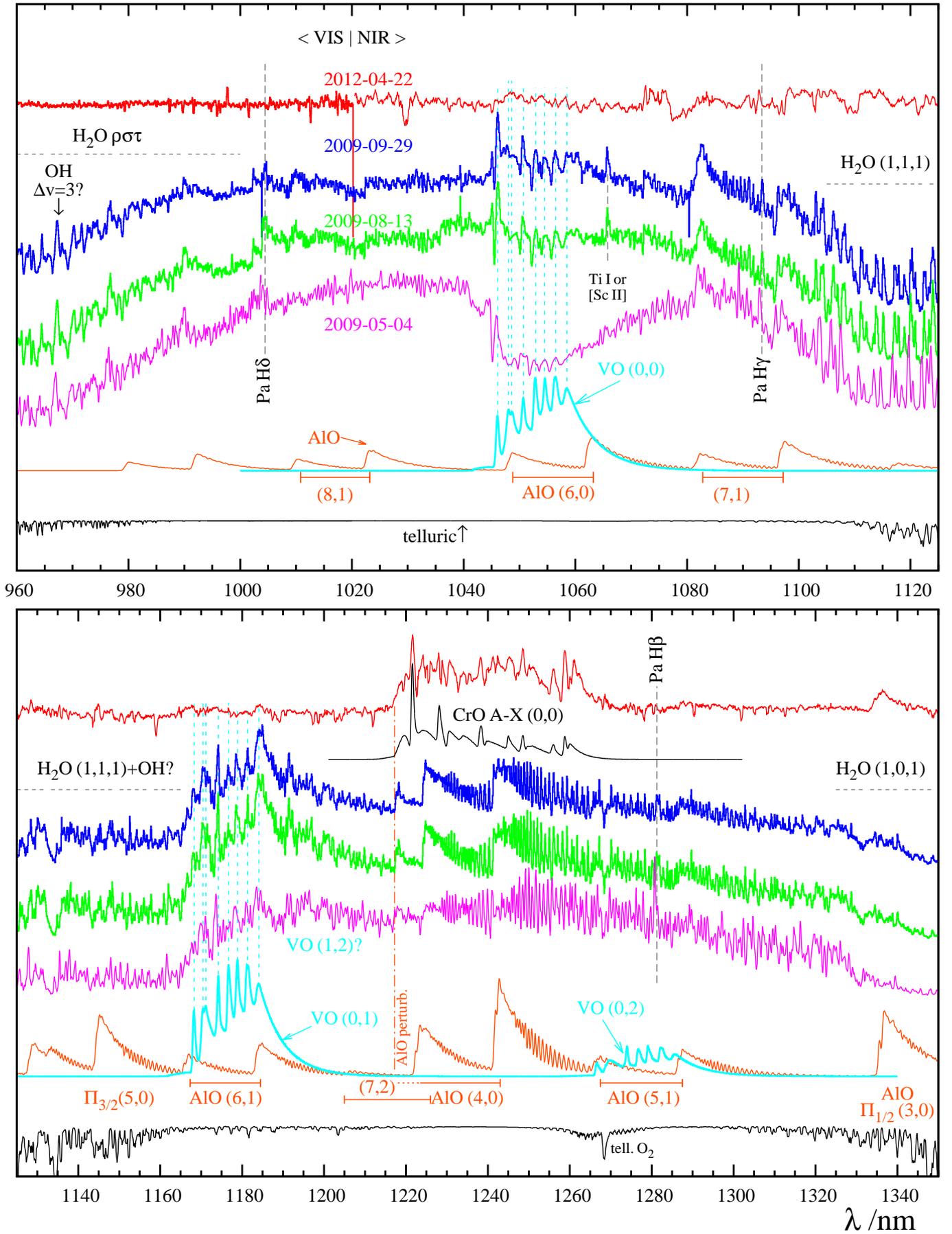}
\caption{NIR spectra of V1309\,Sco covering the $JH$ bands. Simulated spectra of AlO and VO are shown to indicate the location of the emission bands of those species. Location of the water absorption bands is marked with horizontal bars near the edges of the spectral ranges shown. A telluric spectrum is also included at the bottom of the figure (with photospheric absorption features of the standard star). Spectra are in arbitrary units and are not in scale with respect to each other. 
}
\label{fig-nir}
\end{figure*}

\begin{figure*}\centering
\includegraphics[width=\textwidth]{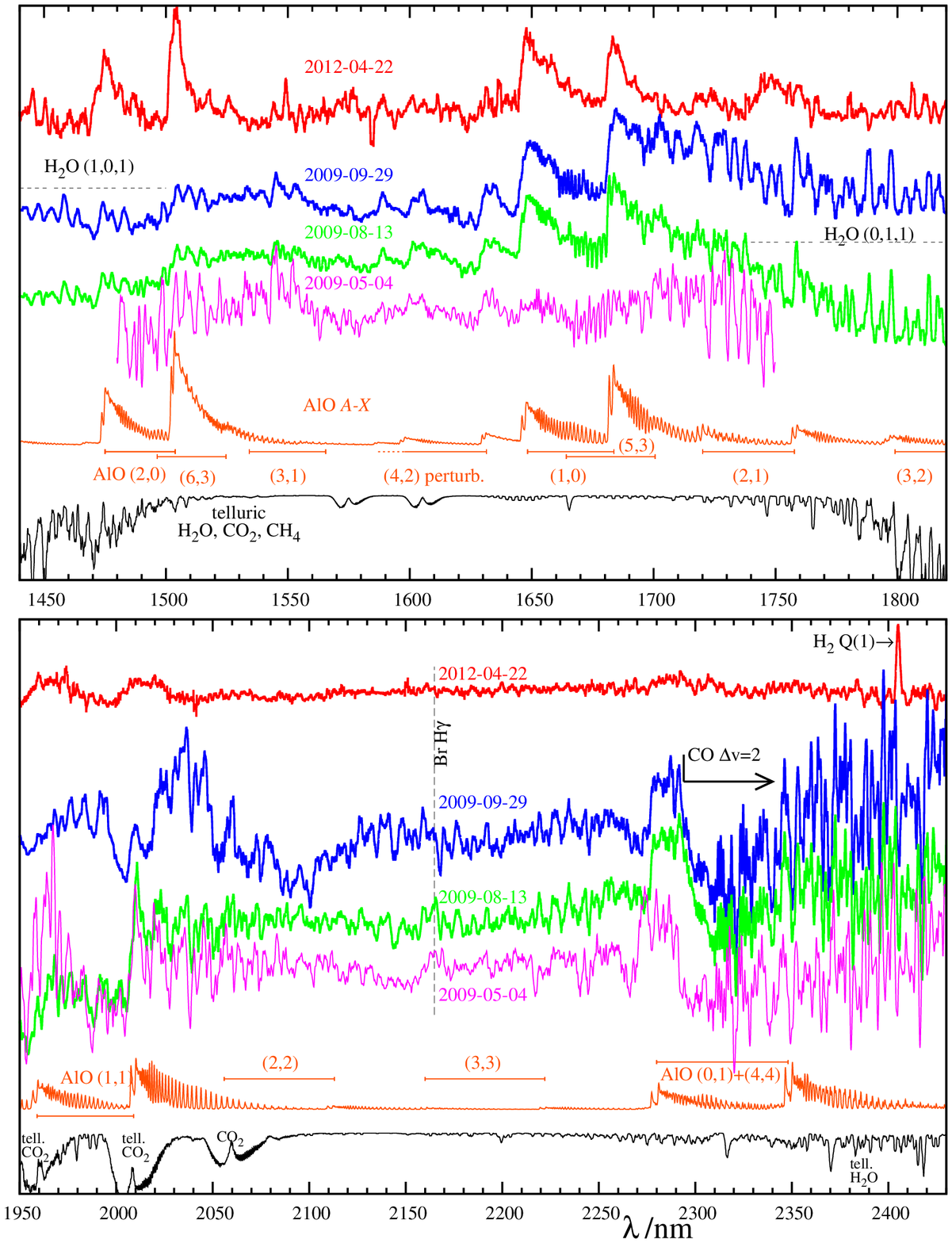}
\caption{The same as Fig.\,\ref{fig-nir} but for spectra covering approximately the $HK$ bands. In addition to AlO, for which a simulation is shown, the first-overtone band of CO is clearly present in the 2009 spectra and one rovib line of H$_2$ is seen in the 2012 spectrum.}\label{fig-nir-B}
\end{figure*}

In the NIR spectra from 2009, emission bands of the $A\,^2\Pi$--$X\,^2\Sigma$ system of AlO were easiest to identify owing to their characteristic double structure and triangular shape of each of the spin components ($\Pi_{1/2}$ and $\Pi_{3/2}$). These infrared bands were first recognized in spectra of V4332\,Sgr in \citet{baner03}. To better characterize the location of all the AlO bands in V1309\,Sco, we performed a simple simulation of the $A-X$ system which is included in Figs.\,\ref{fig-nir} and \ref{fig-nir-B}. Particularly prominent in the covered range are the (4,0) band\footnote{Because our simulation does not include perturbations of the $X\,^2\Sigma\;\varv^{\prime}$=4 state by $A\,^2\Pi\;\varv$=9 \citep{LJ94}, bands involving $\varv^{\prime}$=4 are slightly shifted and incorrect in intensity.}, seen in all spectra and located longward of about 1222\,nm, and   the (1,0) band, located longward of about 1645\,nm;  the emission spike near 2010\,nm, although affected by a (telluric) feature of CO$_2$, is also dominated by the (1,1) band of AlO. Other weaker features are present in the spectra, as can be seen by comparing the observations to the simulation in Figs.\,\ref{fig-nir} and \ref{fig-nir-B}.    



From the band shapes and the relative intensities of the $\Pi_{1/2}$ and $\Pi_{3/2}$ components, it is evident that optical thickness close to the band-heads is very high, i.e. $\tau\gg1$, especially in the $\Pi_{3/2}$ components. Our simulation does not account for such strong saturation effects. The presence of bands involving high vibrational states (at least up to $\varv$=6) and  relative intensities of the bands suggest a high vibrational temperature of 1000--3000\,K, while the band shapes away from band-heads strongly suggest a rotational temperature of a few hundred K. These temperature constraints are  very uncertain, however,  and should be verified by a more realistic excitation model with a proper treatment of optical thickness effects and going beyond the assumption of isothermal gas.

In our latest NIR spectrum from 2012, only the (1,0) and (2,0) bands of $\Pi_{3/2}$, and (3,0) of $\Pi_{1/2}$ are undoubtedly present. The spectral region that was dominated in 2009 by the (4,0) band of AlO evolved until 2012 into a broad feature, which-- as we discuss below-- is dominated by CrO emission. 

\subsubsection{VO}
Also present in the NIR spectra, most clearly in the data from 2009, is the $A\,^4\Pi$--$X\,^4\Sigma^-$ system of VO. The (0,1) band gives rise to an emission feature between about 1165 nm and 1200\,nm that is recognizable by its characteristic eight separate peaks (two features in each quartet component), as illustrated by a simple simulation of the system in Fig.\,\ref{fig-nir}. The band overlaps with a much weaker emission of AlO in band (6,1). The red shoulder of the (0,1) feature of VO may be strongly contaminated by the next band of the $\Delta\varv$=1 progression, i.e. the (1,2) band, which is also expected to have eight peaks. Owing to the lack of appropriate spectroscopic constants, we were not able to include this higher band into our simulation. Only very weak emission of the (0,0) band and perhaps also of (0,2) are present in the 2009 spectra. 

The VO bands are also responsible for absorption features, most clearly in the (0,0) band in the spectrum from May 2009 (commissioning). Most other bands of VO can be interpreted as a combination of emission and absorption features. This also explains why the (0,0) band emission is weaker than in (0,1) in all the 2009 spectra. It is apparent that each of the eight sub-heads of the (0,0) band seen in emission has a corresponding absorption component blue-shifted by about 200\,\kms. This self-absorption profile implies that photons produced by the warm molecular gas responsible for the emission passes through a colder layer of gas. The emission band (0,0) of VO is weak -- yet still recognizable -- in the 2012 spectrum. The time sequence of the NIR spectra of VO illustrates the same changes from pure absorption profiles to emission, similar to what is seen in the visual electronic system of VO.

\subsubsection{CrO}
Using a simulation of the CrO $(0,0)$ band of the $A\,^5\Sigma-X\,^5\Pi$ system, implemented in {\tt pgopher} by C. M. Western, we found a very satisfactory match of the 1220--1270\,nm feature seen in 2012 to the simulated band shape at rotational temperatures close to about 300\,K. The quintet nature of this electronic band is responsible for the broad and very characteristic shape;  we note that the observed profile may be still somewhat contaminated by the bands (4,0) and (5,1) of AlO. This band is remarkably strong and dominates the emission flux in the whole spectrum covered by Xshooter, which is surprising given that this is the first astrophysical identification of this band of CrO. We extended the simulation of the CrO $A-X$ system to include the corresponding (1,0) and the (0,1) bands. Although covered by our 2012 spectrum, the bands fall within absorption bands of water and could not be unambiguously identified. The bands may be weak owing to a low gas vibrational temperature in 2012.

To our knowledge, this is the first astrophysical identification of this infrared CrO band in an astronomical spectrum.


\subsubsection{CO}
The 2009 spectra above 2.3\,$\mu$m\ show the characteristic rotational pattern of the rovibrational (2,0) band of CO (in the ground electronic state $X\,^1\Sigma$). We interpret all the features above 2.3\,$\mu$m as \emph{emission} of the $\Delta\varv$=2 sequence, although some contribution from underlying CO absorption cannot be excluded. The shape of the profile suggests the head of the (2,0) $R$ branch is not formed (or is highly saturated), which implies a rotational temperature below about 1000\,K. However, rotational components from vibrational levels above $\varv$=2, i.e. from (3,1), (4,2), etc., are seen and suggest a much higher vibrational temperature of at least 2000\,K. At the same time, however, there is no evidence for the presence of higher CO overtones in the 2009 spectra. By cross-correlating the observed spectra against our simulation of the $\Delta\varv$=2 sequence, we found the velocity of the two spectra did not change within 2.4\,\kms\ in the 47 days between their exposures and that the mean radial velocity was $-117\pm7$\,\kms\ (in the heliocentric rest frame). There is no significant deceleration of the flow.

In the 2012 spectrum, only weak absorption features can be seen at the expected positions of the main band-head of the $\Delta\varv$=2 sequence of CO and the presence of CO features is tentative at best. If present, the absorption features may originate from a spectrum of one of the blending stars because the location of the absorption bands corresponds to the band-heads  formed at high rotational temperatures, typically met in photospheres of spectral types later than $\sim$G0. The lack of circumstellar CO features in the late spectrum of V1309\,Sco is surprising given the presence of strong and clear emission of rare and exotic species.

\subsubsection{H$_2$}
In the 2012 spectrum where we expected to see CO bands, we observe one emission feature which we identify as the $Q(1)$ (i.e. $J=1-1$) $\varv=1-0$ rovibrational transition of H$_2$ (laboratory wavelength 2406.6\,nm). There are no other features that can be unambiguously identified as H$_2$ transitions. If this identification is correct, all other $\varv=1-0$ lines belonging to the $S$ branch and those in the $Q$ branch from rotational levels $J>1$ must be at least $\sim$6 times weaker than $Q(1)$. In particular, we do not observe the $1-0$ $S(1)$ (i.e. $J=3-1$) line at a laboratory wavelength of 2121.8\,nm which is expected to be of comparable intensity to $Q(1)$ in excitation conditions typical for most known H$_2$ emission sources, i.e. in thermal equilibrium conditions, or fluorescent- or shock-induced excitation \citep[e.g.][]{H2excit}. Perhaps, the line is absent owing to an excitation mechanism that is not efficient above $\varv$=1 $J$=1 making other lines  very weak. The H$_2$ emission extends over a range from --350 to --10\,\kms\ and has a nearly rectangular profile.

\subsubsection{H$_2$O}
Similar to what is seen in the optical region, the NIR spectra show circumstellar features of the rovibrational bands of water in  strong absorption. In fact, the water bands dominate the overall appearance of the NIR spectra in 2009. The lines are broad and blueshifted leaving no doubt they are of circumstellar origin. A simple simulation of the ($v_1, v_2, v_3$)=(1,1,1) band near 1130\,nm -- which (i) has the highest-energy above the ground (6120\,K) among all the observed NIR bands, (ii) is least contaminated by residuals of the telluric correction, and (iii) is least saturated -- gave us an excitation (rotational) temperature of about 1000\,K, i.e. consistent with that derived for the optical $\rho\sigma\tau$ band. The absorption bands to lower-lying vibrational levels, i.e. (1,0,1) and (0,1,1), are saturated and not suitable for a simple excitation analysis as ours. As in the VIS spectrum, there is no sign of water features in the 2012 data, most likely owing to the lack of the continuum source.


\subsubsection{Unidentified features}

Numerous features observed in the NIR spectra remain unidentified. The vast majority of them are broad and display rotational patterns (i.e. a high-frequency ``wave'') that strongly suggests that their origin is molecular. By performing simple simulations of rovibrational or electronic spectra of numerous oxides (MgO $A^1\Pi-X ^1\Pi$, CaO $A'\,^1\Pi-X\,^1\Sigma^+$, and TiO $\phi$ and $\delta$ systems), hydrates (e.g. FeH $X\,^4\Delta-F\,^4\Delta$ and HF $X\,^1\Sigma^+-X\,^1\Sigma^+$), and other compounds (e.g. ZrS $a\,^3\Delta-b'\,^3\Pi$ and CN $X\,^2\Sigma^+-X\,^2\Sigma^+$), we excluded some systems of these species as carriers of the observed features. Unfortunately, some of the bumps seen in the spectra are due to imperfect data reduction and instrumental artefacts whose deforming influence is often very difficult to asses in the NIR part of our Xshooter spectra. 

\section{Line profiles and velocities}
\begin{figure}\centering
\includegraphics[width=\columnwidth]{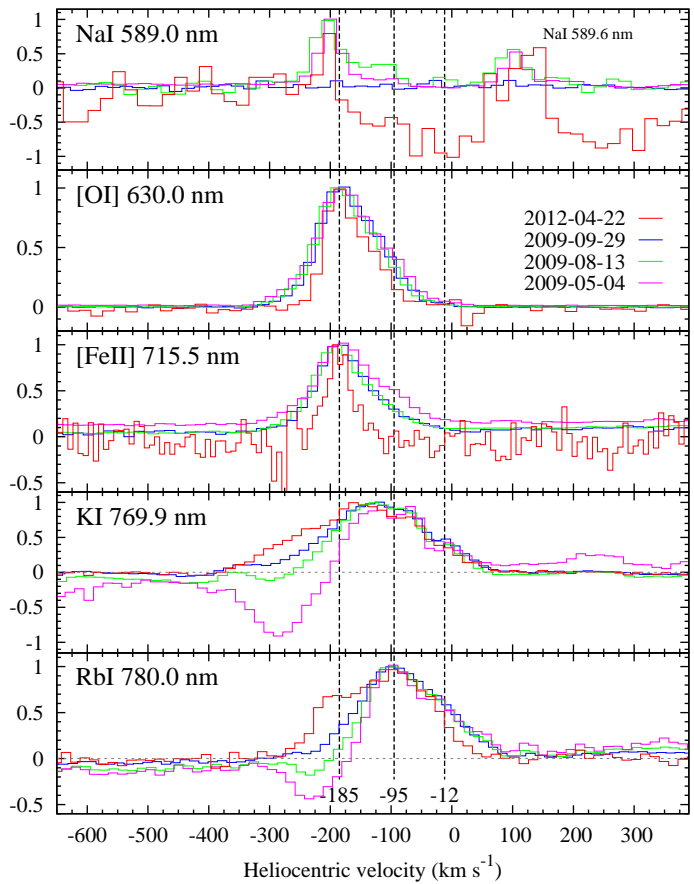}
\caption{Examples of normalized profiles of atomic features as observed at multiple dates in V1309\,Sco. Most characteristic velocities are marked with vertical dashed lines.}\label{fig-profiles}
\end{figure}

The kinematical changes that the remnant underwent during the period covered by our observations can be traced most clearly in the profiles of atomic features. There are two major groups of atomic lines that show distinct type of profiles. Example profiles are shown in Fig.\,\ref{fig-profiles}.

Forbidden emission lines of metals, e.g. \ion{Mg}{I}], [\ion{Fe}{II}],  [\ion{O}{I}], and the Balmer lines, show pure emission profiles over the 2009-2012 period. As described earlier, the lines weaken and become narrower with time. For instance, the 715.5\,nm line of [\ion{Fe}{II}] had a FWHM of 130\,\kms\ in May 2009, 110\,\kms\ in Aug./Sept. 2009, and 57\,\kms\ in 2012; similarly, the width of the strongest line of [\ion{O}{I}] dropped from 131, through 118, to 87\,\kms. The [\ion{O}{I}] and \ion{H}{I} lines also became  more asymmetric with time. These lines are centred at around --180\,\kms\ with a scatter and systematic differences between the different species of a few \kms. In this group, the Balmer lines are  special. The 2009 profiles of H$\alpha$ contain broad Lorentzian-like wings that can be traced as far as $\sim$500\,\kms\ from the line centres, most likely produced as scattering on electrons. The H$\alpha$ profile in 2012 does not show such wings but instead has an extra emission component at $\sim$--12\,\kms. 

Another group of characteristic lines consists of strong resonance doublets of alkali metals, i.e. of \ion{K}{I}, \ion{Rb}{I}, \ion{Cs}{I}, which all showed P-Cyg type profiles before 2012. The absorption component of those profiles was most pronounced in the earliest observations and was becoming weaker with time; in 2012 the absorption component was weak or absent in some of the profiles. The lines of \ion{Cs}{I} diminished until they completely disappeared in 2012. The broadest in this group are lines of \ion{K}{I} and their absorption  extends down to at least $V_h$=--366\,\kms. The emission profiles peak near --95\,\kms\ and at least some of the profiles have multiple components, for instance one peaking at about --12\,\kms. The turn-over point from emission to absorption, which usually indicates the systemic velocity (the kinematical centre  of an outflow), is between --190 and --155\,\kms, depending on the line and the somewhat arbitrary definition of local continuum level. Then the furthest-reaching absorption seen in \ion{K}{I} indicates an outflow terminal velocity of about 200\,\kms. 

Among the lines of alkali metals the profiles of the \ion{Na}{I} optical doublet clearly stand out. The emission is very weak at all epochs. The 2012 spectrum has clear absorption features blueshifted with respect to the faint emission components.  As noted earlier, these features may be a composition of the photospheric and interstellar absorption features in the spectra of the blending stars. In earlier epochs, the interstellar component could have reduced  the strength of the doublet somewhat, but the lines are intrinsically weak. 

The H$_2$ line, which shows pure emission profile, extends over the whole velocity range covered by all the other lines and is clearly different from all the lines described so far. It may contain several discrete emission components blending into this single broad profile. 

Considering the overall kinematical appearance of all observed atomic features, the most characteristic velocity -- which may well be the systemic velocity -- is at about $V_{\rm sys,h}=-185$\,\kms. \cite{mason} derived the systemic velocity of --80$\pm$15\,\kms\ from the position of [\ion{Ca}{II}] lines in their spectra from earlier dates than those considered here. Their value corresponds well to the position of emission components in our spectra, which -- affected by blueshifted absorption --  cannot be a good measure of the systemic velocity. The corresponding LSR velocity of $V_{\rm sys,LSR}=-175$\,\kms\ places V1309\,Sco at the edge of the velocity range seen in the interstellar emission of \ion{H}{I} towards its position \citep{LAB}. This would suggest that V1309\,Sco is a very distant object, perhaps at the outskirts of the Galaxy ($\sim$20\,kpc). However, the distance determinations considered in \citet{tylenda} strongly suggest otherwise,  giving the best estimate of the distance of only 3.5\,kpc. From this, we conclude that the line profiles in V130\,Sco do not provide the true systemic velocity when methods used for classical winds are applied. The extreme negative shift may be perhaps related to a geometry of the outflow which gives us a good view on the approaching parts of the outflow, hiding the redshifted parts.    

The lines do not show any obvious features that would indicate they arise in a disk, existence of which is evident in other observations \citep[][see also below]{tylenda}. For the epochs covered by our data, we also do not favour the model proposed in \citet{mason} of an expanding shell thicker in the equatorial plane and seen from the polar direction. 

\section{Discussion and conclusions}\label{sect-discuss}
\subsection{Temporal changes}
The picture of the evolution of V1309\,Sco arising from our spectral observations is the following. In 2009, the stellar remnant was surrounded by cool molecular gas, but atomic and ionized gas, which dominated the spectra of V1309\,Sco in earlier epochs \citep{mason}, was still present. As our simple modelling of water bands showed, the molecular gas had a rotational (kinetic) temperature lower than about 1000\,K, but the gas was clearly not in thermal equilibrium as even higher vibrational temperatures of up to 3000\,K are evident in spectra of other species. The considerable strengths of emission features shows that the emitting area had to be significantly larger than the stellar disk \citep[e.g.][]{pcyg}. The molecular material was most likely expanding in a form of a shell during whole 2009, as can be seen  by the absorption features becoming weaker with time and by the increasing intensity of emission features (relative to the continuum and the corresponding absorption features); a wind-like outflow is also evident in the P-Cygni type profiles of some atomic lines and molecular features. 

During the whole period covered by our data we clearly observed the Balmer lines and forbidden lines of metals of low ionization degree. This is evidence that hot gas of a temperature of a few thousand K and low density was present in the system. The decreasing flux of these features is due to quick cooling. It is most natural to assume that this is the material heated during the 2008 eruption and still dissipating its energy. However, the observation of the forbidden line of H$_2$ present only in 2012 suggests that the remnant might still have been dynamically active as late as  2012 and produced portions of gas of high excitation. The emission in [\ion{O}{I}] was optically thin during the whole period covered by our observations, as it was in the earlier epochs \citep{mason}.

The source of optical/NIR continuum disappeared sometime between the end of 2009 and April 2012 and the remnant of V1309\,Sco was seen only in discrete spectral emission features in our latest observations. The disappearance may be related to the decrease in brightness of the stellar remnant owing to its cooling but it is likely to be partially related to increased obscuration by dust. 

Dust could already have been  present in V1309\,Sco in 2008 just before the eruption \citep{tylenda,mccollum,zhu}, but  absent before 2008 \citep{nicholls}. As explained in the earlier studies, the dusty material formed sometime before the eruption and is likely confined to a disk seen almost edge on. The optical emission in continuum and lines seen in the 2009 spectra can be interpreted as scattered light of a stellar source that had been considerably obscured to us by the dusty disk. The scattering would occur on the dust (continuum) and associated atomic/molecular gas (lines), just as described in our studies of V4332\,Sgr \citep{kami4332,specpol,tylUVES}. This scattering effect is even more pronounced in V1309\,Sco with a drastically low intensity ratio of the \ion{Na}{I} to \ion{K}{I} doublets.

The fact that we do not see the continuum source in 2012 must be related either to the formation of fresh dust that would increase the extinction towards the source or to some kind of geometrical reconfiguration of the material around the star, for instance by developing a thicker flared disk that would block the light along our line of sight. Infrared photometry from 2010 suggests the former scenario because it shows a sudden increase of reddening of the source \citep{mccollum,nicholls}. Our spectral observations therefore bracket the (second) dust formation period of V1309\,Sco. We note however that, as suggested in \citet{mccollum}, the spectral features in the NIR spectra are strong compared to continuum and can influence interpretation of the broadband photometry in the sense that the fluxes do not correspond to pure continuum, especially after 2009. 



\subsection{V1309\,Sco as a member of the red transient class}
As described in detail in \citet{mason}, the spectral behaviour of V1309\,Sco before 2009 followed that of another red transient, V838\,Mon, which erupted in 2002. The 2009 optical spectra presented here still bear strong similarity to those of V838\,Mon in later stages of its evolution \citep{lynch1,lynch2}, in particular owing to the deep absorption bands of various oxides. The infrared bands of AlO, which are rare in astronomical spectra \citep{baner12} but strong in V1309\,Sco in 2009, were also seen in V838\,Mon, although only in absorption (Lynch et al.).  

The spectral appearance of V1309\,Sco in 2012 is remarkably similar to what has been observed in V4332\,Sgr years to decades after its outburst (1994). In fact, the emission spectrum seen at optical wavelengths is almost exactly the same as that of V4332\,Sgr described in \citet{tylcrause}, \citet{kami4332}, and \citet{tylUVES}. The only difference is the presence of the forbidden emission lines of [\ion{O}{I}] and [\ion{Fe}{II}] in V1309\,Sco, which, we predict, should have disappeared by now (2015). Similarly, we predict that the UVB spectrum of V1309\,Sco will evolve to what is seen in V4332\,Sgr, i.e. a spectrum dominated by low-excitation lines of neutral atoms and molecules. The otherwise very rare infrared features of AlO and CrO seen in emission make late V1309\,Sco most akin to V4332\,Sgr,  as does the silicate dust feature observed near 10\,$\mu$m in both objects \citep[][and references therein]{nicholls}. The geometry and excitation conditions in the merger remnants of both objects are comparable as well.  


The similarity of late spectra of V1309\,Sco to those of the other longer-known red transients is striking. The resemblance strengthens the mutual link between the older red transients and V1309\,Sco and suggests similar physical mechanisms driving their post-outburst evolution. Because the merger event is best seen in V1309\,Sco  (the Rosetta stone) -- owing to a wealth of photometric data before and during the spiralling-in process -- the link provides strong support to the interpretations in which all the older Galactic red transients were indeed caused by stellar mergers.


\subsection{Chemical composition}
The omnipresence of oxides -- in particular the strong water bands seen in the 2009 spectra -- are evidence that the molecular remnant of V1309\,Sco is oxygen rich.
This is in agreement with the presence of inorganic (silicate) dust in the remnant as observed by \cite{nicholls}. There are, however, some noticeable chemical/spectroscopic peculiarities in V1309\,Sco. One is the presence of rare NIR bands of AlO. Their presence in red transients and a relation to formation of alumina dust have been discussed in \citet{baner03,baner26AlO,baner12}. 
 
The presence of CrO in V1309\,Sco (and V4332\,Sgr) is even more remarkable. No other circumstellar spectral features of CrO have been detected so far, including visual, infrared, and millimetre/submillimetre wavelengths \citep{CrO}. The fact that we observe CrO in red transients but not in other known classes of objects implies that either explosive chemistry of red transients favours formation of this particular metal oxide or that CrO is exceptionally abundant owing to an increased abundance of elemental chromium (or only one of its isotopes, see below). Deriving molecular abundances in red transients is very challenging and may in fact be impossible until a very good knowledge of physical conditions and geometry of their circumstellar surroundings is acquired. Even then, deriving elemental abundances from molecular ones would be a very difficult task, especially since these eruptive objects are unlikely to be in chemical equilibrium. Although the field of non-equilibrium chemistry is starting to achieve its first successes \citep[e.g.][]{cherchneff}, much effort is needed before realistic chemical models for objects like V1309\,Sco will be possible.  

Nevertheless, the unique presence of CrO in red transients points to the virtual possibility of abundance anomalies in their remnants. An effort should be taken to identify the remaining unassigned molecular features in the NIR spectrum of V1309\,Sco to investigate whether the potential enhancement in chromium is associated with overabundance of other rare species. The iron-group elements are of particular interest here since oxides of other transition metals (TiO, VO, ScO) are clearly present in V1309\,Sco. Interestingly, it was recently realized that anomalous abundances of $^{54}$Cr, $^{58}$Fe, and $^{64}$Ni in solar system materials may not originate from supernovae, as earlier assumed, but can be a product of $s$-processing in low-mass stars \citep[e.g.][and references therein]{wasserburg}. Some collisions of white dwarfs are also expected to produce material enhanced in iron-group elements, especially in $^{44}$Ti and $^{48}$Cr \citep[e.g.][and references therein]{papish}. Whether the strong signature of CrO in spectra of red novae is a manifestation of a related nucleosynthesis phenomenon is an intriguing question,   especially  in the context of the recent discovery of rare isotopologues in the oldest  recognized red transient, CK\,Vul \citep{CK}.



%
\begin{appendix}
\section{Simulations of the molecular bands}\label{appen}
For identification purposes and first excitation analysis of the observed bands, we performed simple simulations of their electronic or vibrational bands. More detailed excitation analysis for chosen species will be presented elsewhere. All molecules except for water (see below) were performed in {\tt pgopher} \citep{pgopher}. The simulation is very basic and does not include full radiative-transfer calculations. In particular, only a very limited correction for line saturation effect in the high-opacity regime was included. All simulations were performed as for a stationary isothermal layer (slab) of gas. Each species was simulated individually. For some electronic systems, the relative intensities of the different rovibrational transitions were calculated using Franck-Condon factors, if available in the literature. 


\paragraph{AlO $A\,^2\Pi-X\,^2\Sigma$ (NIR):}
Spectroscopic constants were taken from \citet{LJ94} (their Table\,1) and the simulation included vibrational states $\varv^{\prime}\leqslant4$ in the lower electronic state and $\varv^{\prime\prime}\leqslant8$ in the upper electronic state. The Frank-Condon factors (FCFs) were taken from Table\,3 of \citet{BG84} to obtain the relative intensity scale of the bands. Only the main isotopologue, $^{27}$Al$^{16}$O, was included in the simulation; no significant contribution from the unstable Al isotopes was noticed in the observed spectra.

\paragraph{VO $B\,^4\Pi-X\,^4\Sigma$ (VIS):} 
The (visual) $B-X$ system of VO was simulated with spectroscopic constants from \citet{VO1} and \citet{VO2}. Those sources provided us with data for the $\varv$=0 and 1 in both upper and lower electronic states. Some constants were extrapolated to the $\varv^{\prime\prime}$=2 state, allowing for a rough identification of the (0,2) band position but resulting in high uncertainty in the band structure. FCFs of the (0,0) and (1,0) bands were taken from \citet{FCF-VO}. For (0,1) band, the FCF was assumed to be the same as for (1,0), serving only as a very rough approximation on the band intensity. 

\paragraph{VO $A\,^4\Pi-X\,^4\Sigma$ and $D\,^4\Delta-A\,^4\Pi$ (NIR):}
In the NIR region, the electronic structure of VO is very complex, in particular because of overlaps between bands of different systems \citep[see e.g.][]{VO3}. We simulated the NIR spectrum of VO including the (0,0), (0,1), (0,2) bands of the $A-X$ system and the (0,0) band of the $D-A$ system. In the laboratory spectrum of VO, bands with negative values of $\Delta\varv$ in the $B-X$ system (described in the paragraph above) overlap with the near-infrared systems; e.g. the $B-X$ (0,3) band blends with $A-X$ (0,0). We did not include those $B-X$ bands from higher vibrational levels in our simulation, mainly because they are expected to be weak and because of a  lack of spectroscopic constants for the high vibrational levels. For the bands of the $A-X$ system, the spectroscopic constants from \citet{VO4} were used.  Our {\tt pgopher} simulation of this system also  included  the relative band intensities based on FCFs from \citet{FCF-VO}. We also included the (0,0) band of the $D-A$ system, which would overlap partially with the $A-X$ (0,0) band, but found no indication of its presence in the observed spectra. The simulation was made only for the main isotopologue, $^{51}$V$^{16}$O. 

\paragraph{CrO $A\,^5\Sigma-X\,^5\Pi$ (NIR):}
The $A-X$ (0,0) band of CrO was simulated in {\tt pgopger} using an example file provided by the program's author, C. Western, and based on spectroscopic constants from \citet{CrOCheung}. Using spectroscopic data of \citet{Barnes93} for the $\varv=1$ level of the $^5\Pi$ electronic state, we extended the simulation to the (0,1) band, which is predicted to be located near 1365--1430\,nm. This places the band within the saturated band of water and therefore the observed spectra could not be conclusively tested for the presence of this CrO band.  We also added the (1,0) band by calculating spectroscopic data for the $\varv=1$ state at $A\,^5\Sigma$ from the $\Delta G_{1/2}$ value from \citet{CrOCheung}. This allowed us to predict the origin of this band near 1100\,nm, a region dominated by absorption of water. The position of the strongest head (spike) within the band, however, overlaps with a discrete emission-like feature seen in all 2009 spectra, but the shape and the central position are not so well matched as in the case of the (0,0) band. We therefore could not conclusively decide on its presence.

\paragraph{CrO $A^{\prime}\,^5\Delta-X\,^5\Pi$ (VIS):}
The band $A^{\prime}-X$ (0,0) of CrO was simulated using spectroscopic constants from \citet{Barnes93} provided for the $\varv=0$ and 1 states within the upper and lower electronic levels. For $\varv=1$ at $X\,^5\Delta$, the band origins are given individually for the five different components arising from the spin-orbit interaction. We calculated the corresponding vales parametrizing the interaction (i.e. $T_0, A, \lambda, \eta, \theta$, fixing $\gamma$ at 0.00), following formula (19) in \citet{CrOCheung}. With these constants, we performed simulations for the (1,0), (0,0), (1,1), and (0,1) bands. However, so far the laboratory spectra have only been  reported for the overlapping (0,0) and (1,1) bands so our simulation of the (1,0) and (0,1) bands is rather uncertain. The observed shape of the (0,0) band suggests a rotational temperature of lower than $\sim$300\,K and may be as low as a few tens of K.

The predicted position of the (0,1) band near 905\,nm overlaps with the strong $\rho\sigma\tau$ band of water and the (0,2) band of VO hampering any conclusive identification. If the vibrational temperature is close to the rotational temperature we are able to roughly constrain, this band should be very weak. The (1,0) should be stronger. The predicted location of the (1,0) band overlaps with the strong 0,0 band of VO at about 790\,nm. There are several strong features with no identification within the VO band whose shape is similar to that of the simulated (1,0) band of CrO. However, the position of the simulated band misses  the unidentified emission by $\sim$1.6\,nm. Better spectroscopic data for vibrationally excited levels are needed to conclusively decide whether the observed emission can be identified as CrO bands.

\paragraph{H$_2$O (VIS/NIR):}
Simulations of water bands were based on a line list of \citet{barber} and included only the main isotopologue, H$_2^{16}$O. Unlike for all the other molecules, the simulations of H$_2$O were obtained with the {\tt codespectra-BT2.f90} program, which is distributed with the line list. The line  intensities are calculated under conditions of thermal equilibrium and within the limit of  optically thin lines. The spectra were simulated as arising from an isothermal and stationary slab of material.

\end{appendix}

\begin{acknowledgements}
Based on observations made with the Very Large Telescope at Paranal Observatory under programmes 60.A-9445(A), 60.A-9024(A), and 089.D-0041(A). We thank Prof. A. Udalski and the OGLE team for providing us with the photometric data of V1309\,Sco. 
\end{acknowledgements}

\end{document}